\documentclass[aps,prd,onecolumn, notitlepage,superscriptaddress,10pt,nofootinbib,twocolumn,
amsmath,amssymb]{revtex4-1}

\usepackage{color}
\usepackage{graphicx}
\usepackage{lipsum}  
 
\newcommand{\fett}[1]{\boldsymbol{#1}}

\newcommand{\dd}{{\rm{d}}}

\newcommand{\be}{\begin{equation}}
\newcommand{\ee}{\end{equation}}

\newcommand{\nabq}{\fett{\nabla}_{\fett{q}}}
\newcommand{\nabx}{\fett{\nabla}_{\fett{x}}}

\definecolor{darkred}{rgb}{0.5,0,0}
\definecolor{darkgreen}{rgb}{0,0.5,0}
\definecolor{darkblue}{rgb}{0,0,0.5}

\usepackage{hyperref}
\hypersetup{colorlinks,
linkcolor=darkblue,
filecolor=darkgreen,
urlcolor=darkred,
citecolor=darkblue}

\begin{document}

%\preprint{\vbox{\hbox{\hfil AEI-2014-002}}}

\title{Frame dragging and Eulerian frames in general relativity}
%\title{Frame dragging from the perspective of a relativistic Lagrangian perturbation theory}

\date{\today}

\author{Cornelius Rampf}
\email{cornelius.rampf@aei.mpg.de}
\affiliation{Max-Planck-Institute for Gravitational Physics (Albert-Einstein-Institute), D--14476 Potsdam-Golm, Germany}
\affiliation{School of Physics, University of New South Wales, Sydney, New South Wales 2052, Australia}

\begin{abstract}
The physical interpretation of %the gravitational evolution of irrotational 
cold dark matter perturbations is clarified by associating Bertschinger's Poisson gauge with a 
Eulerian frame of reference. We obtain such an association by using a Lagrangian approach to relativistic cosmological structure formation. 
Explicitly, we begin with the second-order solution of the Einstein equations in a synchronous/comoving  coordinate system, which defines the Lagrangian frame, and transform it to a Poissonian coordinate system. 
The generating vector of this coordinate/gauge transformation is found to be the relativistic displacement field.
The metric perturbations in the Poissonian coordinate system contain known results from standard/Eulerian Newtonian perturbation theory, but  contain also purely relativistic corrections. On subhorizon scales, these
relativistic corrections are dominated by the Newtonian bulk part. 
These corrections, however, set up nonlinear (initial) constraints for the density and for the velocity that become important on scales close to the horizon.
Furthermore, we report the occurrence of a transverse component in the displacement field, and  find that it induces a nonlinear frame dragging as seen in the
 Eulerian frame, which is subdominant at late times and subhorizon scales.
Finally, we find two other gauges that can be associated with a Eulerian frame. We argue that the Poisson gauge is to be preferred because it comes with the simplest physical interpretation.

\end{abstract}

\maketitle   

%%%%%%%%%%%%%%%%%%%%%%%%%%%%%%%%%%%%%%%%%%%%%%%%%%%%%%%%%%

\section{Introduction}

It is believed that the large-scale structure of the Universe is the result of gravitational instability.\footnote{Report number: AEI-2014-002.}
The governing evolution equations are provided by general relativity (GR), 
although the simpler Newtonian theory yields reasonable estimates at most scales of interest.
Exact analytic solutions, for generic initial conditions and without any symmetry, 
in both GR and the Newton theory are not possible, so one has to use either numerical approaches 
(Newtonian $N$-body simulations) or analytical approximations (cosmological perturbation theory \cite{Bardeen:1980kt,Kodama:1985bj,Mukhanov:1990me,Ma:1995ey,Bruni:2003hm,Malik:2008im,Bernardeau:2001qr}; 
CPT). In CPT, the equations of motion for cold dark matter (CDM) are usually solved within the 
irrotational-fluid-dust approximation, which restricts the validity of the approach to 
sufficiently large scales. 
The (additional) use of the Newtonian approximation, on the other hand, is  
assumed to be valid only on interaction scales well below the causality bound.
To study the evolution of perturbations close to the causality bound, a relativistic treatment becomes mandatory.

We should seek for a relativistic treatment accompanied with a direct correspondence to the Newtonian solutions.
Only such a treatment is capable to deliver straightforward physical interpretations, since one can
parametrise the relativistic corrections as deviations from the Newtonian bulk part. 
A close correspondence becomes 
increasingly important especially when studying "gauge-dependent'' (frame-dependent) quantities as we shall do in the following. 

Lagrangian perturbation theory (LPT) is a promising avenue of the gravitational instability, 
mostly since it is an intrinsically nonlinear approach to nonlinear structure formation 
 but also as it is required to set up initial conditions for $N$-body simulations. 
Additionally, the Lagrangian representation comes with a simple physical interpretation
as one follows simply the trajectories of fluid elements.
The only dynamical quantity in Newtonian LPT is the displacement field $\fett{F}$, which parametrises the gravitationally induced deviation of the fluid element from its initial Lagrangian position $\fett{q}$.
 The Newtonian coordinate transformation to the Eulerian  coordinate $\fett{x}$ is
\be
 \label{newtCo} \fett{x}(t,\fett{q})= \fett{q}+\fett{F}(t,\fett{q}) \,.
\ee
The Newtonian LPT has inspired hundreds of works since Refs.~\cite{Zeldovich:1969sb,Buchert:1989xx,Buchert:1992ya,Ehlers:1996wg}. 
Explicit solutions up to third order in LPT were derived in Refs.~\cite{Buchert:1993ud,Bouchet:1994xp}. 
The fourth-order scheme in the LPT was derived in Ref.~\cite{BuchertRampf:2012}, 
and a general recursion relation in LPT was reported in Ref.~\cite{Rampf:2012up}. 
Important improvements  about the LPT related to convergence issues were recently given in Refs.~\cite{NadkarniGhosh:2010th,NadkarniGhosh:2012gm}.

Significant efforts have been made to obtain a general relativistic generalisation of LPT; see, e.g.,~\cite{Matarrese:1994wa,Russ:1995eu,Buchert:2012mb,Buchert:2013qma}. 
Recently, we obtained a relativistic generalisation of LPT \cite{Rampf:2012pu,Rampf:2013ewa} from a somewhat different perspective than the aforementioned references; we identified LPT in terms 
of a coordinate transformation of a perturbed synchronous metric, resulting from a relativistic gradient expansion, %---resulting from
% solving the Einstein equations for an irrotational dust component up to second order, 
transformed to a Eulerian/Newtonian coordinate system. This perspective offers a unique interpretation of 
gauge transformations in GR, a perspective we shall further develop in the following. Furthermore, by including not only scalar perturbations but also vector and tensor perturbations, we generalise the findings of Refs.~\cite{Rampf:2012pu,Rampf:2013ewa}.

Identifying and interpreting relativistic effects within cosmological structure formation
%---and interpreting these relativistic effects  within Newtonian $N$-body simulations, 
are the two key objectives we shall study in this paper. We specifically focus on 
 relativistic effects of the density and the velocity field as (would be) measured from (relativistic) $N$-body simulations. 
% than extracted from real observations, where the latter would be flawed by secondary distortions (such as coming from the photon propagation). 
To understand this paper it is very helpful to recall that the density and velocity fields are \emph{Eulerian fields}, 
although corresponding counterparts could be defined in any other frame.
%, e.g., one can define a fluid density in the Lagrangian frame. 
%In Newtonian \emph{and} relativistic cosmology, the density and velocity fields are frame dependent. 
Especially when considering observables
(e.g., correlators of the density/velocity field) from $N$-body simulations only the Eulerian density and the Eulerian 
velocity, evaluated at the Eulerian position, are the objects of interest.\footnote{Here and in the following we neglect geometrical and dynamical distortions coming from the propagation of the photons in a clumpy and expanding Universe. We also neglect biasing effects in this {paper}. We thus assume that the density/velocity perturbations are directly \emph{observable}. This view is obviously not realistic for  cosmological observations, but this perspective is necessary to investigate in frame/gauge dependence as we shall do in the following. Note, however, that the density/velocity perturbations are indeed directly observable in (relativistic) $N$-body simulations.}
%Thus our investigation is not philosophical but rather essential, e.g., to understand $N$-body simulations in terms of GR.}
For example, as is well known from results of the Newtonian perturbation theory, only $N$-point correlators (or their counterparts in Fourier space) 
from the Eulerian density/velocity field yield reasonable approximations especially when compared with results from $N$-body simulations
(e.g., \cite{Bernardeau:2001qr,Matsubara:2007wj,Rampf:2012xb}).
%\footnote{Note that at leading order we have for the Newtonian $N$-point correlators of the density in the Eulerian and Lagrangian frame: $\langle \rho(\fett{x}_1) \cdots \rho(\fett{x}_N) \rangle \equiv \langle \rho(\fett{q}_1) \cdots \rho(\fett{q}_N) \rangle$, and similarly for their counterparts in Fourier space. Thus the disparity between the Eulerian and Lagrangian frame appears only beyond leading order.}
Here, it is but one essential task to show that we can draw  similar conclusions for relativistic cosmologies---keeping in mind, however, that we do not yet have access to fully relativistic $N$-body simulations (but see \cite{Adamek:2013wja}), so an explicit verification is not yet feasible.

That frame dependence affects not only Newtonian cosmologies but also relativistic ones was recently shown in Refs.~\cite{Rampf:2012xb,Rampf:2013ewa,BruniHidalgo2013}. 
In these, the authors derived the density contrast $\delta \equiv (\rho-\overline \rho)/\overline \rho$ up to second order with the use of the CPT or related techniques. 
The easiest way to make the frame dependence explicit is to focus on the Newtonian part of their relativistic density contrast. According to Refs.~\cite{Rampf:2012xb,Rampf:2013ewa,BruniHidalgo2013}, the (spatial part of the) density contrast in the synchronous/comoving gauge reads at second order $\delta_{\rm Lagrangian}^{(2)}(\fett{q}) \propto F_2 - \Phi_{,l}\Phi_{,lmm}$ plus relativistic corrections, where the Newtonian bulk part $F_2$ is given in Eq.\,(\ref{F2Grad}), 
%and the notation will be explained in detail in the following section.
 and $\Phi$ is the primordial potential. 
By transforming the density contrast to "some'' Eulerian frame, however, we have at second order $\delta_{\rm Eulerian}^{(2)} (\fett{x}) \propto F_2$ plus some relativistic corrections.  Apart from the relativistic corrections, precisely the same happens in the Newtonian Eulerian/Lagrangian correspondence, and the disapperance of the additional term in $\delta_{\rm Lagrangian}^{(2)}$ is very well understood (see also the Appendix).\footnote{Depending on the explicit (experimental) setup, however, also $\delta_{\rm Lagrangian}(\fett{q})$ could become relevant in relativistic cosmology.}

The interpretation of the density and velocity field is inherently linked with the 
proper identification of the Eulerian frame.
In the Newtonian approximation the identification of the Eulerian frame is trivial, and the connection to the Lagrangian frame is given by the coordinate transformation~(\ref{newtCo}). This identification is however nontrivial in the relativistic generalisation (see section~\ref{other?}); there is generally no preferred coordinate system in GR, and as a consequence there is no single frame which could be labelled as Eulerian. As we shall see in the following, in GR there exists a class of coordinate systems which can be associated with a Eulerian frame. The essential idea here is to use the Newtonian correspondence from LPT to identify "a'' Eulerian frame in GR, preferably 
a Eulerian frame  (which turns out to be the one associated with the Poisson gauge) accompanied with simple physical interpretations.
Thus, fairly analogous to the Newtonian coordinate transformation~(\ref{newtCo}), we define its relativistic counterpart to be
\be
   {x}^\mu(t,\fett{q}) = {q}^\mu + F^\mu(t,\fett{q}) \,, \label{GRtrafo}
\ee 
where $\mu$ are the four space-time components (since it is the four-dimensional line element that is 
invariant in GR);  $x^\mu \dot= (\tau,\fett{x})$ and $q^\mu \dot= (t,\fett{q})$ are the Eulerian and Lagrangian coordinates, respectively, and $F^\mu \dot= (L,\fett{F})$ is the relativistic displacement field. 
Thus, the displacement field now consists not only of a spatial  but also of a temporal part. This 
is nothing but the statement that space and time are on an equal footing in GR. Physically it means that space and time will mix due to the nonlinear clustering. The coordinate transformation~(\ref{GRtrafo}) is the central building block to formulate a relativistic LPT. Our procedure to obtain a relativistic LPT can be summarised as follows: 
\begin{enumerate}
 \item[(1.)] Find a relativistic solution in a synchronous/comoving coordinate system. 
 \item[(2.)] Identify the corresponding frame to be Lagrangian.
 \item[(3.)] Use Eq.\,(\ref{GRtrafo}) to find $F^\mu$ and the metric perturbations in the  
   "new'' coordinate system with coordinates $x^\mu$. 
 \item[(4.)] Identify the very coordinate system to be a Eulerian frame, if the metric potentials and the displacement field agree with Newtonian results (at least) in the weak-field limit.
\end{enumerate}
We shall use this procedure to find all Eulerian frames. Note that essentially the last point in the above list defines what we mean as a Eulerian frame in GR.
We call these coordinate systems "Eulerian'' since they yield the correct Newtonian bulk part of the density and velocity.

One important application of this paper is certainly related to generating initial conditions and about the interpretation of $N$-body simulations from the perspective of GR.
Some investigations have been made about Newtonian $N$-body simulations and their compatibility with GR \cite{Flender:2012nq,Haugg:2012ng}. An explicit recipe to interpret $N$-body results with respect to GR at linear order in the CPT was first given in Refs.~\cite{Chisari:2011iq,Green:2011wc}. It is also known that GR yields an initial constraint 
for the density field beyond leading order \cite{Bartolo:2010rw,Hidalgo:2013mba,Rampf:2013ewa,BruniHidalgo2013}, although its interpretation and practical implementation are still in their beginnings \cite{Rampf:2013ewa,Rigopoulos:2013nda}. Here, we seek to gain further understanding of this issue. Moreover, we report the occurrence of an additional nonlinear  
constraint coming from GR, which affects the velocity field---already at initial time. Specifically, we obtain
a nonzero transverse component in the Lagrangian displacement field that is the result of the nonlinear coordinate transformation. In the Eulerian frame this phenomenon appears as a non-linear frame dragging. 

In general, the occurence of a non-zero transverse component in the relativistic Lagrangian displacement field is expected to happen at some order in perturbation theory, even within the restrictive class of an irrotational fluid velocity---a 
restriction we also consider here. 
Indeed, similar considerations within the Newtonian limit of the LPT with equivalent initial conditions
were studied in detail (e.g., Refs.~\cite{Buchert:1993ud,Bouchet:1994xp,Catelan:1994ze}), 
and a nonzero transverse displacement field  was found at third order in Newtonian LPT. This transverse displacement field  can be interpreted as a fictitious force, very similar to the Coriolis force, induced through a noninertial motion of the fluid element \cite{Catelan:1994ze}. The transverse displacement field therefore corrects the motion of the fluid element, and it is thus essential to include it in the analysis---neglecting it would formally lead to wrong results.
%\footnote{The neglection of the transverse displacement field thus yields a violation of the irrotationality condition, i.e., in the Newtonian limit $\fett{\nabla_x} \times \fett{u} \neq \fett{0}$, where $\fett{u}$ is the particle velocity.} as has been shown in the Newtonian analysis of Ref.~\cite{Rampf:2012xb}. 
Similar conclusions can be made for the general relativistic treatment. 
%In particular, the longitudinal and transverse part of the relativistic displacement field can be used to set up quasi-relativistic $N$-body simulations.
%

This paper is organised as follows. In Sec.~\ref{sec:grad} we review the metric up to second order in a synchronous/comoving coordinate system, which was obtained in Refs.~\cite{Russ:1995eu,Rampf:2012pu,Rampf:2013ewa}. 
This metric serves as the starting point for the current investigation and will define the Lagrangian frame. 
%Explicitly, the synchronous/comoving metric contains not only scalar perturbations but also vector and tensor perturbations which are excited from scalar perturbations at 
%the linear level.
By the coordinate transformation to a Poissonian coordinate system, described in Sec.~\ref{sec:GCT}, we shall obtain the Lagrangian displacement field and a physical interpretation of the perturbations in the Eulerian frame.
In Sec.~\ref{sec:iter} we explain how to solve the coordinate transformation with an iterative technique, and we also define useful operators that are needed in the latter. Then, we report the first-order and second-order results of the transformation in Secs.~\ref{sec:sol1} and \ref{sec:sol2}, respectively, and discuss them in detail. 
In Sec.~\ref{other?}  we report a procedure to identify all possible Eulerian gauges.
Explicitly, we find three (nontrivial) gauge choices that can be associated with a Eulerian frame, and 
we clarify their physical interpretations.
 We summarise and conclude afterwards in Sec.~\ref{sec:con}. We also wish to highlight the Appendix in which we relate our findings to the Newtonian approximation.

For simplicity, we restrict our current investigation to 
a Universe with only a CDM component (thus we set the cosmological constant $\Lambda$ to zero). 
Our results hold for a $\Lambda$CDM Universe to a fairly good approximation,
 since the reported relativistic corrections become important only on very large scales where the cosmological constant should not have much influence \cite{Rampf:2012pu}.

\quad

\quad

%%%%%%%%%%%%%%%%%%%%%%%%%%%%%%%%%%%%%%%%%%%%%%%%%%%%%%%%%%%%%%%

\section{Metric perturbations in a synchronous coordinate system}\label{sec:grad}

In this section we report the relativistic solution in a synchronous/comoving coordinate system. 
 Here we only introduce our conventions and report the final result for an irrotational CDM component up to second order; explicit calculations  can be found in, e.g.,~Ref.~\cite{Matarrese:1994wa} and, in particular, Ref.~\cite{Russ:1995eu} for the tetrad formalism and Refs.~\cite{Rampf:2012pu,Rampf:2013ewa} for the gradient expansion technique.

The corresponding comoving/synchronous line element is
\be
  \dd s^2 = - \dd t^2 + a^2(t)\,\gamma_{ij}(t,\fett{q}) \, \dd q^i \dd q^j \,,
\ee
where $t$ is the proper time of the fluid element and $\fett{q}$ are 
comoving/Lagrangian coordinates, constant for each
pressureless and irrotational fluid element; in this paper we assume an Einstein-de Sitter (EdS) Universe with the cosmological scale factor $a(t) = (t/t_0)^{2/3}$.
Summation over repeated indices is implied---for Latin indices from 1 to 3 and for Greek indices from 0 to 3. We define this coordinate system to be the Lagrangian frame. This definition is possible and unique, and the spatial part of the synchronous coordinate $\fett{q}=const.$ fixes the initial position for the fluid element (CDM particle).
Inflation predicts at linear order the initial seed metric
\be
  k_{ij} =  \delta_{ij} \left[ 1+ \frac{10}{3} \Phi(\fett{q}) \right]  \,,
\ee
where $\Phi(\fett{q})$ is the primordial Newtonian potential, given at initial time 
$t_0$. 
In our case $\Phi(\fett{q})$ is just a Gaussian field, and it is directly related to Bardeen's gauge-invariant 
potential \cite{Bardeen:1980kt}. 
Here and in the following, a "$,i$'' denotes a differentiation 
with respect to Lagrangian coordinate $q_i$. 

Solving the Einstein equations with the use of the initial seed metric $k_{ij}$ and some iterative technique,
we obtain for an EdS universe up to second order 
\begin{widetext}
\begin{align} 
  \gamma_{ij} (t,\fett{q}) &=  \, \delta_{ij} \left( 1+ \frac{10}{3} \Phi \right)  
  + 3a(t)t_0^2 \left[ \Phi_{,ij} \left( 1-\frac{10}{3} \Phi \right) 
      - 5 \Phi_{,i} \Phi_{,j} +  \frac 56 \delta_{ij}  \Phi_{,l} \Phi_{,l} \right] \nonumber \\
   &\quad  -\left( \frac{3}{2}  \right)^2 \frac 3 7 a^2(t) t_0^4 
   \Bigg[ 4\Phi_{,ll} \Phi_{,ij}   \Bigg.  
  -  \delta_{ij} \left( \Phi_{,ll} \Phi_{,mm} - \Phi_{,lm}\Phi_{,lm}  \right)   \Bigg] 
   + \left( \frac{3}{2} \right)^2 \frac{19}{7} a^2(t)t_0^4 \,\Phi_{,li} \Phi_{,lj} + \chi_{ij}(t,\fett{q})  \,, \label{gammaUniverseGrad}
\end{align}
\end{widetext}

where we have retained only the fastest growing mode solutions (see Ref.~\cite{Rampf:2013ewa} for the inclusion of decaying modes). The divergence-free and trace-free tensor  $\chi_{ij}(t,\fett{q})$  is of order $\Phi^2$,  and it results from the magnetic part of the Weyl tensor---its explicit form is not needed in the following, but see, e.g.,~\cite{Matarrese:1994wa,Russ:1995eu}. Note that $\chi_{ij}$ is not determined by the gradient expansion of Refs.\,\cite{Stewart:1994wq,Rampf:2012pu,Rampf:2013ewa}. The inclusion of $\chi_{ij}$ in the following does not change our conclusions, and we just include it for the sake of generality.

%%%%%%%%%%%%%%%%%%%%%%%%%%%%%%%%%%%%%%%%%%%%%%%%%%%%%%%%%%
\section{Coordinate transformation}\label{sec:GCT}

To obtain the relativistic displacement field 
% and the perturbations in the Eulerian frame 
we perform a coordinate transformation to the Poisson gauge. We transform  
the result\,(\ref{gammaUniverseGrad}) written in the synchronous/comoving gauge with coordinates $(t,\fett{q})$,
\begin{widetext}
\begin{align}
  \dd s^2  &= g_{\mu\nu}(t,\fett{q}) \,\dd q^\mu \dd q^\nu \,
   =  - \dd t^2 +a^2(t)\gamma_{ij} (t,\fett{q}) \,\dd q^i \dd q^j \,, 
\intertext{to the Poisson gauge with coordinates $(\tau, \fett{x})$ and corresponding metric ($\tau$ is \emph{not} the conformal time)}
  \dd s^2 &=  g_{\tilde\mu\tilde\nu}(\tau,\fett{x}) \,\dd x^{\tilde\mu} \dd x^{\tilde\nu}            = - \big[ 1 \big. +2 \big. A(\tau,\fett{x}) \big] \dd \tau^2  + 2a(\tau) w_i(\tau,\fett{x}) \, \dd \tau \dd x^i  + a^2(\tau)  \left\{ \left[ 1- 2 B(\tau,\fett{x}) \right] \delta_{ij} + S_{ij}(\tau,\fett{x}) \right\} \dd x^i \dd x^j \,.
\end{align}
\end{widetext}
$A$, $B$, $w_i$, and $S_{ij}$ are supposed to be small perturbations.  The tensor $S_{ij}$ is traceless, 
i.e.~${S^i}_i = 0$. The Poisson gauge is defined via \cite{Matarrese:1997ay,Ma:1995ey,Bruni:2013mua}
\begin{align}
 \begin{split}
   \partial^{x_i} w_i &=0 \,, \\ 
   \qquad 
 \partial^{x_i} S_{ij}  &= \partial^{x_i} S_{ij}^{\rm T} = 0 \,,
 \end{split} \qquad 
  \text{(gauge conditions).} \label{gauge}
\end{align}
These conditions hold also in the perturbative sense.
The two coordinate systems are related by the coordinate transformation
\begin{align}  
 \label{LtrafoGrad}
   &x^\mu(t,\fett{q}) = q^\mu +  F^\mu (t,\fett{q}) \,,
\intertext{with}  
    &x^\mu = \begin{pmatrix} \tau \\ \fett{x} \end{pmatrix}, \qquad q^\mu = \begin{pmatrix} t \\ \fett{q} \end{pmatrix} \,, % \qquad  \text{and} 
  \qquad F^\mu =  \begin{pmatrix} {L} \\ \fett{F} \end{pmatrix} \,,
\end{align}
where 
${L}(t,\fett{q})$ and $\fett{F}(t,\fett{q})$ are supposed to be small perturbations.  $\fett{F}$ is the spatial part of the 
relativistic Lagrangian displacement field, and %(cf.~with the Newtonian coordinate transformation, Eq.\,(\ref{newtCo})). 
$L$ is the time perturbation---in the case of the Poisson gauge, $L$ is the velocity potential of the fluid element (i.e., this is generally not true for other Eulerian gauges; see Sec.~\ref{other?}). Note explicitly that $L$ contains only the potential part of the velocity of the fluid element; 
thus the full 3-velocity field is given by the time derivative of the 3-displacement field, i.e., $\fett{u}= a\,\partial \fett{F}/\partial t$.
We decompose $\fett{F}$ into a curl-free and divergence-free vector field,
\be
\fett{F}(t,\fett{q}) = \fett{F}^\parallel(t,\fett{q})
 +\fett{F}^\perp(t,\fett{q}) \,, 
\ee
and without loss of generality we choose to decompose it with respect to the Lagrangian coordinate system.

General covariance requires the invariance of the line element $\dd s^2$, and thus 
\be \label{gct}
  g_{\mu \nu}(t,\fett{q}) = \frac{\partial x^{\tilde\mu}}{\partial q^\mu} \frac{\partial x^{\tilde\nu}}{\partial q^\nu}
    g_{\tilde\mu \tilde\nu} (\tau,\fett{x}) \,.
\ee
We  shall solve the above general coordinate transformation perturbatively, whilst expanding all fields and dependences.

%%%%%%%%%%%%%%%%%%%%%%%%%%%%%%%%%%%%%%%%
\section{Iterative solution scheme and useful projection operators}\label{sec:iter}

The general coordinate transformation~(\ref{gct}) gives separate equations for the space-space, space-time and time-time parts, which can be used to constrain the parameters $(A,B,w_i,S_{ij},L,F_i)$. We solve these equations order by order. Formally, each small 
quantity is expanded in a series, i.e.,
\begin{align}
A &= \epsilon A^{(1)} +\epsilon^2 A^{(2)}  +\ldots \,, \nonumber \\ 
B &= \epsilon B^{(1)} +\epsilon^2 B^{(2)}  +\ldots \,, \nonumber 
\end{align}
etc., where $\epsilon$ is supposed to be a small dimensionless parameter. The primordial potential $\Phi$ is of order $\epsilon$. 
For convenience we truncate the coordinate transformation of the metrics, Eq.\,(\ref{gct}), up to second order and suppress the perturbation parameter $\epsilon$ in the following. After some manipulations, we find for Eq.\,(\ref{gct})
\begin{widetext}
\begin{align}
  \gamma_{ij}(t,\fett{q}) \simeq  &-\frac{L_{,i} L_{,j}}{a^2} 
    +2 \frac{L_{,(i}  w_{j)}}{a}  +\delta_{ij} \left[ 1 -2B(\tau,\fett{x})  + \frac{4L(t,\fett{q})}{3t} +\frac{2L^2}{9t^2} - \frac{8 B L}{3t} \right]  \nonumber \\
     & +2F_{(i,j)} (t,\fett{q})\left( 1-2B + \frac{4L}{3t}\right) 
\label{trunc1}    + F_{l,i} F_{l,j} +S_{ij}(\tau,\fett{x}) +2 F_{l,(i}  S_{j)l} \,, \\
 0 \simeq &- \left(  1+2A  + \frac{\partial L}{\partial t} \right) L_{,i} 
      +a^2(t) \left[ 1-2B+\frac{4L}{3t} \right] \frac{\partial F_i(t,\fett{q})}{\partial t} +a^2 S_{im} \frac{\partial F_m}{\partial t}
  + a^2 F_{l,i} \frac{\partial F_{l}}{\partial t} \nonumber \\
\label{trunc2}  &  +a(t) \,w_i(\tau,\fett{x}) \left[ 1+\frac{2L}{3t} + \frac{\partial L}{\partial t} \right] + a w_l F_{l,i}
    \,,\\
 -1 \simeq &-1 -2A(\tau,\fett{x}) -2 \frac{\partial L(t,\fett{q})}{\partial t}
  -4 A \frac{\partial L}{\partial t} -\left( \frac{\partial L}{\partial t} \right)^2
  +2a\, w_l \frac{\partial F_l}{\partial t}
 \label{trunc3}  +a^2 \frac{\partial F_l}{\partial t} \frac{\partial F_l}{\partial t} \,.
\end{align}
\end{widetext}
We have suppressed some dependences where there is no confusion; i.e., dependences in
second-order terms can be interchanged, and the resulting error is only of third order.

We solve Eqs.\,(\ref{trunc1})--(\ref{trunc3}) with an iterative technique. 
For that purpose, the decomposition valid for any tensor $T_{ij}$ is useful \cite{Bertschinger:1993xt},
\begin{align}
 T_{ij} = \frac{\delta_{ij}}{3} \hat Q + \left( \partial_i \partial_j  -\frac{\delta_{ij}}{3} \fett{\nabla}^2  \right) \hat T^\parallel + 2 \hat T_{(i,j)}^\perp + \hat T_{ij}^{\rm T} \,,
\end{align}
where $\hat Q$ is the trace of $T_{ij}$; $\hat T_i^\perp$ is a divergence-free vector; and for the transverse traceless tensor, we have $\partial^j \hat T_{ij}^{\rm T}=0$.
It is then straightforward to define the corresponding projection operators 
\begin{align}
 \label{operators} \begin{split}
   \hat T^\parallel &= \frac 3 2 \frac{\partial^i \partial^j}{\fett{\nabla}^2 \fett{\nabla}^2} T_{ij} 
             - \frac 1 2 \frac{1}{\fett{\nabla}^2} \hat Q \,, \\
 \varepsilon^{kli} \hat T_{i,l}^\perp &= \frac{1}{\fett{\nabla}^2}  \varepsilon^{kli} \partial^j \partial^l T_{ij} \,,
 \end{split}
\end{align}
where $\varepsilon^{kli}$ is the Levi-Civit\`a symbol, and $1/\fett{\nabla}^2$ is the inverse Laplacian.

With the above we can extract the relevant information from Eqs.~(\ref{trunc1})--(\ref{trunc3})
 to obtain the following:
\begin{enumerate}
 \item The Lagrangian displacement field $\fett{F} = \fett{F}^\parallel+\fett{F}^\perp$.
   The operator $\hat T^\parallel$ applied to Eq.\,(\ref{trunc1}) 
   constrains the longitudinal part of the displacement field $\fett{F}^\parallel$, 
   whereas $\varepsilon^{kli} \hat T_{i,l}^\perp$ 
   constrains its transverse part $\fett{F}^\perp$. 
 \item The divergence of Eq.\,(\ref{trunc2}) constrains the time perturbation $L$.
 \item The scalar perturbation $B$ is obtained by the trace part of Eq.\,(\ref{trunc1}).
 \item The curl of Eq.\,(\ref{trunc2}) constrains the vector perturbation $\fett{w}$.
 \item From Eq.\,(\ref{trunc3}) we obtain the scalar perturbation $A$.
\end{enumerate}
In the following we solve the coordinate transformation with that procedure, order by order. 

Before proceeding, it is worthwhile to compare this procedure with other methods in the literature.
One crucial extension in this procedure compared to the one 
in Refs.~\cite{Rampf:2012pu,Rampf:2013ewa} is the consistent inclusion of the transverse displacement field $\fett{F}^\perp$ in the coordinate transformation. 
$\fett{F}^\perp$ has to be included since $\partial^i S_{ij}\equiv0$, but the same is \emph{generally not true} for the divergence of Eq.\,(\ref{trunc1}). Thus, $\fett{F}^\perp$ absorbs the transverse part of Eq.\,(\ref{trunc1}), and relates it to $\fett{w}$ via Eq.\,(\ref{trunc2}). Indeed, the following identity is valid at least up to second order: 
$\fett{w}  \equiv - a \,\partial \fett{F}^\perp / \partial t$. In Refs.\,\cite{Rampf:2012pu,Rampf:2013ewa} the coordinate transformation is performed from the synchronous gauge to the Newtonian gauge---instead of the Poisson gauge, which is the generalisation of the Newtonian gauge. In the Newtonian gauge vector and tensor perturbations are set 
to zero by hand; thus $\fett{w} := \fett{0}$ and so is $\fett{F}^\perp = \fett{0}$. This is, however, rather accidental, and there is generally no reason to discard the transverse component of the displacement field.

%%%%%%%%%%%%%%%%%%%%%%%%%%%%%%%%%%%%%%%%%%%%%%%%%%%%%%%%%
\section{Solutions up to first order in the Poisson gauge}\label{sec:sol1}

 With the use of the above recipe and with the gauge conditions~(\ref{gauge}), we obtain up to first order for Eq.\,(\ref{LtrafoGrad})
\renewcommand*{\arraystretch}{1.4}
\begin{align}
 \label{firstorder_displacement}
  F_\mu^{(1)} (t,\fett{q}) &= \begin{pmatrix} L^{(1)} \\ \fett{F}^{(1)\parallel}+\fett{F}^{(1)\perp} \end{pmatrix} \,,
\end{align}
with
\begin{align}
 \label{firstOrderTrafo}  L^{(1)}(t,\fett{q})   &= \Phi(\fett{q})\,t  \,, \\
  F_i^{(1)\parallel}(t,\fett{q})  &=   \frac 3 2 a(t)t_0^2 \, \partial_{q_i} \Phi(\fett{q}) \,, \\
  F_i^{(1)\perp}(t,\fett{q})  &= 0 \,,
\end{align}
and for the scalar, vector and tensor perturbations, respectively,
\begin{align}
 \label{Poiss1}  \begin{split}
   A^{(1)}(\tau,\fett{x}) &= B^{(1)} (\tau,\fett{x}) = - \Phi(\fett{x}) \,, \\
 \qquad w_i^{(1)}  (\tau,\fett{x})  &= 0 \,, \\
 S_{ij}^{(1)}  (\tau,\fett{x}) &= 0 \,.
 \end{split}
\end{align}
$F_i^{(1)}$ is the displacement field in 
the Zel'dovich approximation \cite{Zeldovich:1969sb}; since it is purely longitudinal the trajectory of the fluid element is just along the overall potential flow.
$L^{(1)}$ is the peculiar velocity potential of the fluid element, and the perturbations
$A^{(1)}$ and $B^{(1)}$ in the Poisson gauge, Eq.\,(\ref{Poiss1}), are in agreement with 
the weak-field limit of general relativity. Thus, we recover the Newtonian approximation at linear order. 
Note that we have interchanged in Eq.~(\ref{Poiss1}) the dependence of $\Phi$ such 
that $\fett{q} \rightarrow \fett{x}$, which is only valid up to first order. At second order, we simply have to Taylor expand the dependence 
$\Phi(\fett{q}) \simeq \Phi(\fett{x}-\fett{F}^{(1)})$, in accordance with the coordinate transformation~(\ref{LtrafoGrad}).

%%%%%%%%%%%%%%%%%%%%%%%%%%%%%%%%%%%%%%%%%%%%%%%%%%%%%%%%
\section{Solutions up to second order in the Poisson gauge}\label{sec:sol2}

Similar considerations can be made up to second order. We obtain the second-order quantities
\begin{align}
 \label{Fmu2} F_\mu^{(2)} (t,\fett{q}) = \begin{pmatrix} L^{(2)} \\ \fett{F}^{(2)\parallel}+\fett{F}^{(2)\perp} \end{pmatrix} \,,
\end{align}
with
\begin{align}
 \label{eqL2} L^{(2)} (t,\fett{q})  &= \frac 3 4 t^{5/3} t_0^{4/3} \Phi_{,l} \Phi_{,l}- \frac 9 7 t^{5/3} t_0^{4/3} \frac{1}{\nabq^2} \mu_2 \nonumber \\
   &\qquad\quad     -\frac 7 6 t \Phi^2 +4t C\,,\\ 
 \label{2LPT} {F}_i^{(2)\parallel} (t,\fett{q}) &=  - \left( \frac 3 2 \right)^2 \frac 3 7 a^2(t)t_0^4 \frac{1}{\nabq^2} \partial_{q_i} \mu_2  \nonumber \\
   &\qquad\quad-5a(t) t_0^2 \partial_{q_i} \Phi^2 + 6a(t)t_0^2  \partial_{q_i} C \,, \\
 \label{2LPTtrans} F_{i}^{(2)\perp} (t,\fett{q})  &=  6a(t)t_0^2  R_i  \,,
\end{align}
and the scalar, vector and tensor perturbations, respectively, up to second order (for convenience, we include the first-order perturbations),
\begin{align} \label{AB}
 \begin{split}
  A(\tau,\fett{x}) &\simeq \phi_{\rm N} - 4 \overline C \,, \qquad      B(\tau,\fett{x}) \simeq \phi_{\rm N} +\frac 8 3 \overline C \,, \\
   w_i(\tau,\fett{x})  &= -4 \tau^{1/3} t_0^{2/3} \overline R_i \,, \qquad S_{ij}^{(2)}(\tau,\fett{x}) = \overline \chi_{ij} \,, 
 \end{split}
\end{align}
with
\begin{widetext}
\begin{align}
  \mu_2 (t,\fett{q}) &\equiv  \frac 1 2 \left( \Phi_{,ll} \Phi_{,mm} - \Phi_{,lm} \Phi_{,lm}  \right) \,, \\
\label{kernelC}  C  (t,\fett{q}) &\equiv \frac{1}{\nabq^2 \nabq^2}  \left[ \frac 3 4 \Phi_{,ll} \Phi_{,mm} + \Phi_{,m} \Phi_{,llm} + \frac 1 4 \Phi_{,lm} \Phi_{,lm}  \right] \,,\\
 \label{kernelR} R_i (t,\fett{q}) &\equiv \frac{1}{\nabq^2 \nabq^2}  \left[  \Phi_{,il}\Phi_{,mml}  - \Phi_{,lli} \Phi_{,mm}
      + \Phi_{,i} \Phi_{,llmm} -\Phi_{,m} \Phi_{,mlli} \right]  \,, \\
 \label{PhiN} \phi_{\rm N} (\tau,\fett{x}) &\equiv - \Phi(\fett{x}) +\frac 3 2 a t_0^2 \frac{1}{\nabx^2} \left[ \frac 5 7 \Phi_{|ll} \Phi_{|mm} + \Phi_{|l} \Phi_{|lmm} + \frac 27 \Phi_{|lm} \Phi_{|lm} \right] \,.
\end{align}
\end{widetext}
The definitions of $\overline C$, $\overline R_i$, etc.~are identical to $C$, $R_i$, etc., but with dependences and derivatives interchanged to
$(\tau,\fett{x})$ rather than $(t,\fett{q})$. We denote spatial derivatives with respect to the Poissonian (Eulerian) coordinate $x_i$ with a slash.
In $A$ and $B$ we have neglected terms proportional to $\Phi^2$ that are not enhanced by spatial gradients.

The first term in  Eq.\,(\ref{2LPT}) contains the second-order improvement from Newtonian LPT, whereas the remnant terms are the same relativistic corrections as in Refs.~\cite{Rampf:2012pu,Rampf:2013ewa}. Equation~(\ref{eqL2}) is the velocity potential of the displacement field.
The transverse part of the displacement field, Eq.\,(\ref{2LPTtrans}), together with the corresponding $\fett{w}$ is one of our main results:
\be
 \label{MaxwellFaraday} \fett{w} = -  a \frac{\partial \fett{F}^\perp}{\partial t} \,.
\ee
This expression (which is generally nonzero already at initial time; see the following section) clearly indicates the gravitomagnetic origin of the frame dragging:
The frame dragging vector potential $\fett{w}$ is directly related to the transverse part of the fluid's velocity.\footnote{Note that the peculiar velocity of the fluid element is $\fett{u} = a\, \partial \fett{x} / \partial t \equiv a\, \partial \fett{F} / \partial t$, where the last step follows from $\fett{x}(t,\fett{q}) = \fett{q} +\fett{F}(t,\fett{q})$.}

Since initial conditions are set within the linear regime, where the transverse displacement field is (at least perturbatively) suppressed, the above shows that the frame dragging will grow as soon as nonlinearities will form. Therefore, the frame dragging gets enhanced by the nonlinearities in the  gravitational evolution. This argument is also valid for the linear frame dragging that we do not consider here since we study only the evolution of irrotational fluids.

Equations~(\ref{AB}) contain the results in the Poissonian coordinate system. The expression $\phi_{\rm N}$, Eq.\,(\ref{PhiN}), matches exactly Newtonian Eulerian perturbation theory (see the Appendix), whereas the remnant terms in $A$ and $B$ denote relativistic corrections that are proportional to the nonlocal kernel $C$, given in Eq.\,(\ref{kernelC}). These results agree with the treatment of Ref.~\cite{Rampf:2013ewa} in the Newtonian gauge and therefore generalises their results to the inclusion of vector and tensor perturbations. 
The occurrence of known results from Newtonian Eulerian perturbation theory in the Poissonian coordinate system indicates that we can associate the Poissonian coordinate system with a Eulerian frame. We shall further specify the labelling "Eulerian'' in section~\ref{other?}, but before that we wish to analyse the consequences (Sec.~\ref{sec:constraint}) and the origin (Sec.~\ref{sec:origin}) of the (transverse) perturbations and also discuss our results with respect to known investigations in the literature (Sec.~\ref{sec:compare}).

%%%%%%%%%%%%%%%%%%%%%%%%%%%%%%%%%%%%%%%%%%%%%%%%%%%%%%%%%%
\subsection{Nonlinear initial constraints for the density and the velocity field}\label{sec:constraint}

In the previous section, we obtained the Lagrangian displacement field $F^\mu$, which  
does not only include the known Newtonian part but also some relativistic corrections. It is very important to note that these relativistic corrections are only partly a result from the gravitational evolution. The other part results from the nonlinear constraints of the velocity field. Crucially, we find that these nonlinear constraints are already apparent at initial time.

It is actually straightforward to understand these initial constraints in terms of the displacement field. To do so, we first have to restore the decaying modes, which we have neglected before because of simplicity. We then obtain for the 4-displacement field $F^\mu \dot= (L,\fett{F})$ up to second order
\begin{align}
 L(t,\fett{q})  &=  v \,\Phi + \left( \frac 3 2 \right)^2 a^2 \Bigg[ \frac{\dot D D}{2} \Phi_{,l}\Phi_{,l}
    + \dot E \frac{1}{\nabq^2} \mu_2 \Bigg. \Bigg.\Bigg] \nonumber \\
  &+ v  \left[ v H +\frac 3 4 a^2 \ddot D -\frac 5 3 \right] \Phi^2 \nonumber \\
 \label{time} & +v  \left[ 2 vH +3a^2 \ddot D + \frac{10}{3}\right] C , \\
 F_i(t,\fett{q}) &= \frac 3 2 D \Phi_{,i}  +\left( \frac 3 2 \right)^2\!E
    \frac{\partial_{q_i}}{\fett{\nabla}_{\fett{q}}^2} \mu_2 \nonumber \\
 \label{spatial} &-5 D \partial_{q_i} \Phi^2 + \left[ 5 D + \left(\frac{v}{a}\right)^2 \right] \left\{ \partial_{q_i} C + R_i \right\} \,,  
\end{align}
where $H$ is the Hubble parameter (here, $H \equiv 2/(3t)$), and we have defined 
\be
v(t) \equiv \frac 3 2 a^2 \dot D \,.
\ee
A dot denotes a partial differentiation w.r.t.~Lagrangian time $t$, and the general growth functions are \cite{Rampf:2013ewa}
\begin{align}
 \label{evoDE}
 \begin{split}
   D(t) &= \frac{20}{9} \int^t \frac{\dd t'}{a^{2}(t')} J(t')  \,, \\
   E(t)  &=  \frac{200}{81} \int^t \frac{\dd t'}{a^2(t')}   \left[ \frac{K(t')}{a^2(t')}  -  \frac{9}{10} D(t') J(t')   \right]  \,,
 \end{split}
\intertext{with}
 \label{evoJK}
 \begin{split}
  J(t) &= \left[2a(t)\right]^{-1} \int^{t}  a(t') \,\dd t' \,,  \\
  K(t) &= a(t)  \int^{t} a^{-1}(t') J^2(t') \, \dd t' \,.
 \end{split}
\end{align}
Generally, one could specify initial data for two coefficient functions out of 
three, namely for the coefficients of the displacement, of the velocity, and of the acceleration. 
To disentangle the effects coming solely from the gravitational evolution (which are unwanted for this demonstration), however, it is useful to require the vanishing of the growth functions at initial time: $D(t_0) = E(t_0) =0$. Now, setting the coefficient functions of the velocity field, $\dot D$ and  $\dot E$, to zero at initial time would be unphysical \cite{Rampf:2013ewa,Buchert:2013qma} (see also the following paragraph). The precise settings for the velocity coefficients do not matter here so we just require that  
$\dot D(t_0) \neq 0$ and $\dot E(t_0) \neq 0$. From that, it follows that $v(t_0) \neq 0$. Again, this means that the fluid element receives an initial nonzero velocity. Thus, the spatial part of the displacement field becomes at initial time
\be \label{NIC}
   \lim_{t \rightarrow t_0} F_i(t,\fett{q}) = v^2(t_0) \left\{ \partial_{q_i} C + R_i \right\} \neq 0 \,.
\ee
This has two important consequences.
 First, the first term is purely longitudinal so it feeds back to the relativistic Poisson equation at initial time: It yields an initial density perturbation \cite{Rampf:2013ewa}. Second, \emph{both} terms in Eq.~(\ref{NIC}) yield nonlinear initial constraints for the velocity field of the fluid element; specifically for its longitudinal and transverse part, respectively. To our knowledge, the latter has not been reported in the literature.

Similarly,  we find that the vector perturbation in the Poisson gauge is initially nonzero, because it is sourced by the initial transverse velocity: $\fett{w}(t_0) =  \partial \fett{F}^\perp(t_0) /\partial t \neq \fett{0}$. Explicitly, violating $\fett{w}(t_0)\neq \fett{0}$ would yield unphysical initial conditions for the CDM fluid element. Another way to understand the situation is the following. Assume that one \emph{could} set the transverse velocity to zero intitially. Then, the vector perturbation $\fett{w}$ would be switched off at initial time and then switched on during the gravitational evolution. This is wrong since the vector perturbation is not  the result of the gravitational evolution but the result of the non-linear structure of Einstein's equations. This argument can be easily verified by studying the time derivative of the transverse part in Eq.~(\ref{spatial}) at initial time.

%%%%%%%%%%%%%%%%%%%%%%%%%%%%%%%%%%%%%%%%%%%%%%%%%%%%%%%%%%
\subsection{Origin of the transverse displacement field}\label{sec:origin}

At second order the spatial coordinate transformation is not entirely longitudinal anymore, i.e., the Lagrangian displacement field acquires a nonzero transverse part. Physically, the transverse displacement field is needed to correct for the actual direction of the fluid motion. Technically, the occurence of a 
transverse displacement field is expected to happen at some order since the coordinate transformation (and thus the fluid's motion as well) is nonlinear and noninertial.

The transverse displacement field is by definition a vector perturbation.
It is important to note that the vector perturbation is not generated through the coordinate transformation itself, but it is already nonzero in the synchronous coordinate system.
To see this, we apply the second operator in Eq.~(\ref{operators}) on the 3-metric $\gamma_{ij}$ (see Eq.\,(\ref{gammaUniverseGrad})), i.e., 
$\varepsilon^{kli} \partial^j \partial^l \gamma_{ij} \equiv  \fett{\nabla}^2 \varepsilon^{kli} \hat\gamma_{i,l}^\perp$. 

%\newpage 

\noindent Then, we find the divergenceless vector
\be
  \hat\gamma_i^\perp = - 5 at_0^2 R_i +\left( \frac{3}{2} \right)^2 a^2t_0^4 Q_i \,,
\ee
with $R_i$ the same as in Eq.\,(\ref{kernelR}), and
\begin{align}
 Q_i &= \frac{1}{\fett{\nabla}^2 \fett{\nabla}^2} \Big[ \Phi_{,lmi} \Phi_{,lmjj} - \Phi_{,lmm}\Phi_{,ljji} \Big. \nonumber \\
   & \qquad\qquad\qquad  \Big. + \Phi_{,li} \Phi_{,lmmjj} - \Phi_{,lm}\Phi_{,lmjji} \Big] \,.
\end{align}
Since $\hat\gamma_i^\perp$ is given in a synchronous/comoving coordinate system, 
where the velocity of the fluid element is by definition zero, only the transformation to a Eulerian frame leads to a physical interpretation of the divergenceless vector.
The physical interpretation is that the transverse displacement field appears as a
 nonlinear frame dragging in the Eulerian frame.

%%%%%%%%%%%%%%%%%%%%%%%%%%%%%%%%%%%%%%%%%%%%%%%%%%%%%%%%%%
\subsection{Comparison of our results with the literature}\label{sec:compare}

Despite different used techniques, similar results were  
reported earlier in the literature \cite{Matarrese:1997ay,Takada:1997bk}. 
In Ref.~\cite{Matarrese:1997ay} (see also Ref.~\cite{Mollerach:2003nq}), authors obtained
the second-order metric perturbations for an EdS Universe in the synchronous/comoving gauge by the use of CPT. They performed a gauge transformation from the synchronous/comoving gauge to the Poisson gauge. 
Although not directly apparent, our results agree where possible (note that they use the conformal time). Explicitly, by translating their time evolution factor $\tau$ according to $\tau \rightarrow 3 a^{1/2}$, and noting that their $\Psi_0 \equiv -(1/\nabla^2) \mu_2$, and their $\Theta_0 \equiv - \frac 2 3 C$, their Eqs.~(6.6)--(6.8) deliver the correct gauge generator and perturbations in the Poisson gauge up to second order, as reported here (apart from factors of $\Phi^2$).\footnote{It is not so straightforward to see e.g.~that $\Theta_0 \equiv - \frac 2 3 C$. See the Appendix for technical details.} However, they did not recognise that this gauge generator is indeed the Lagrangian displacement field, neither that it contains the well-known Newtonian displacement field. Similarly, the time component of the gauge generator, their Eq.\,(6.6), is the Lagrangian velocity potential, but this identification is missing, too.\footnote{Note that in Ref.~\cite{Matarrese:1997ay} the gauge generator is given in terms of the Poissonian spatial coordinate, i.e., $\fett{x}$, whereas we evaluate it at the spatial coordinate of the synchronous/comoving coordinate system, i.e., $\fett{q}$. This explains the occurrence of two additional terms in their Eqs.~(6.6).} Additionally, they do not give initial data for the displacement field and its velocity. So we think that our approach is more suited if it comes to the physical interpretation of the relativistic corrections.

In Ref.~\cite{Takada:1997bk}, the authors developed a post-Newtonian approach to the LPT. 
Their longitudinal post-Newtonian solution for the displacement is given in their
Eq.~(3.66). Unfortunately, due to  the complexity of their expression we cannot confirm whether their longitudinal solution agrees with ours; we leave this issue for some future investigation. For the transverse part of the displacement field, their Eq.~(3.88), we find agreement with our Eq.\,(\ref{2LPTtrans}), if we replace their redefined potential $\fett{\Psi}^{(1)}$ with our potential $\Phi$. However, due to some restrictions of their approach, they were not able to require any initial data for the transverse displacement. So they had to assume \emph{some} initial data, and they set the initial transverse velocity to zero. As thoroughly explained in Sec.~\ref{sec:constraint}, this setting is wrong as it ``switches off'' the vector perturbations at initial time. 

It is important to note that the partial agreement of our results with the ones in Refs.~\cite{Matarrese:1997ay,Takada:1997bk} is highly nontrivial, since the reported results rely on entirely different techniques (i.e., gradient expansion vs.~conventional CPT vs.~post-Newtonian LPT).
So we consider the mutual agreement as a strength in such that our reported relativistic corrections for the longitudinal and transverse displacement field do not depend on the used perturbative scheme. Note that only our approach yields initial non-linear constraints for the density and for the (longitudinal and transverse) velocity field.

%%%%%%%%%%%%%%%%%%%%%%%%%%%%%%%%%%%%%%%%%%%%%%%%%%%%%%%%%%
\section{Are there other Eulerian frames?}\label{other?}

Within the Newtonian approximation there exists only one Eulerian frame. In GR, however, the
situation is generally more complicated, just because there are so many possibilities to choose the coordinate system (i.e., the gauge). One would naturally ask whether other coordinate systems can be identified to be Eulerian (we will also define what we consider as a Eulerian frame). 
Indeed, we will show in the 
following that there are three  Eulerian coordinate systems, but we argue that the Poissonian coordinate system is accompanied with the easiest physical interpretation.
 Thus, the Poissonian coordinate system is a preferred Eulerian frame.

For convenience we restrict to first-order perturbations in the following, 
and we leave a full second-order treatment for future investigations. 
We can then neglect vector and tensor perturbations because we assume them to be of second order.
As before, we define the 
synchronous/comoving coordinate system with coordinates $(t,\fett{q})$ to be associated with the Lagrangian frame ($\fett{q}\equiv const$). The metric perturbations in the Lagrangian frame read
\begin{align}
  \dd s^2 = - \dd t^2  &+ a^2(t) \Big[ \delta_{ij} \left( 1+ \frac{10}{3} \Phi \right) + 3a(t)t_0^2 \Phi_{,ij} \Big] \dd q^i \dd q^j .
\end{align}
Consider the  
first-order coordinate transformation from  {the} unique Lagrangian frame 
to some Eulerian frame
\begin{align}
 \label{GCT}
  x^\mu(t,\fett{q}) &= q^\mu + F^\mu(t,\fett{q})\,,
\end{align}
{with}  
\begin{align}
    &x_\mu = \begin{pmatrix} \tau(t,\fett{q}) \\ \fett{x}(t,\fett{q}) \end{pmatrix}, \quad q_\mu = \begin{pmatrix} t \\ \fett{q} \end{pmatrix} \,, % \qquad  \text{and} 
  \quad F_\mu =  \begin{pmatrix} {L}(t,\fett{q}) \\ \fett{F}(t,\fett{q}) \end{pmatrix} \,,
\end{align}
where the corresponding generic scalar metric of the Eulerian frame, first without any gauge fixing, is
\begin{align}
\dd s^2 &= - \big[ 1 \big. +2 \big. A \big] \dd \tau^2  
 + 2a w_{,i} \, \dd \tau \dd x^i  \nonumber \\ 
 &\qquad\quad + a^2  \left\{ \left[ 1- 2  B \right] \delta_{ij} +2h_{,ij}\right\} \dd x^i \dd x^j\,. 
\end{align}
 We follow Ref.\,\cite{Flender:2012nq} and
calculate the proper time between two events along a worldline, which reads

\newpage

\begin{widetext}
\begin{align}
 \int \sqrt{- \dd s^2} &= \int \dd \tau \sqrt{1+2A 
   -2a w_{,i} \frac{\dd x^i}{\dd \tau}    -a^2 [(1+2 B) \delta_{ij} +2h_{,ij}] \frac{\dd x^i}{\dd \tau} \frac{\dd x^j}{\dd \tau}
  } \nonumber \\
 &\simeq \int \dd \tau \left(  1+A -a w_{,i} \frac{\dd x^i}{\dd \tau}
    -\frac{a^2}{2} \delta_{ij} \frac{\dd x^i}{\dd \tau} \frac{\dd x^j}{\dd \tau} \right)
  \equiv \int \dd \tau \tilde{L} \,.
\end{align}
\end{widetext}
In the last steps we only kept terms of ${\cal O}(g \frac{\dd x^i}{\dd \tau})$, where
$g \in \left\{ A, B, w, h, \dd x^i / \dd \tau\right\}$. The proper time is only extremal if 
the integrand $\tilde L$ satisfies the Euler--Lagrange equation
\be \label{EL}
 \frac{\dd}{\dd \tau} \frac{\partial \tilde L}{\partial \left( \frac{\dd x^i}{\dd \tau} \right)}
  = \frac{\partial \tilde L}{\partial x^i} \,.
\ee
We then obtain up to first order \cite{Flender:2012nq}
\begin{align} 
 \label{no_gauge}
 \frac{\dd}{\dd \tau} \left( a w_{,i} +a^2 \frac{\dd x_i}{\dd \tau}\right) &= -A_{,i} 
   \quad (\text{no~gauge~fixing}).
\intertext{This looks almost like Newton's law of motion which is at first-order}
 \label{Euler} 
 \frac{\dd}{\dd \tau} a^2 \frac{\dd x_i}{\dd \tau} &= \Phi_{,i} 
  \quad \,\,\, \, (\text{Euler equation}).
\end{align}
 Again, we did not yet specify a gauge in Eq.~(\ref{no_gauge}). 
Our aims at this stage are 
\begin{itemize}
\item to find all possible gauges that lead exactly to the Euler equation~(\ref{Euler})
  at linear order,
\item to establish the weak-field limit between the Lagrangian and Eulerian frames.
\end{itemize}
The former implies that we are only interested in trajectories that are Newtonian-like in the weak-field limit.
The latter implies that we have to encode the spatial information of the trajectory in the Zel'dovich displacement field. Thus, {we set} ${F}_i^{(1)}= \frac 3 2 at_0^2 \Phi_{,i}$ for the coordinate transformation~(\ref{GCT}), but leave the temporal perturbation $L$ first unfixed.
Studying the coordinate transformation for the Lagrangian and Eulerian metrics
\be
  g_{\mu \nu}(t,\fett{q}) = \frac{\partial x^{\tilde\mu}}{\partial q^\mu} \frac{\partial x^{\tilde\nu}}{\partial q^\nu}
    g_{\tilde\mu \tilde\nu} (\tau,\fett{x}) \,,
\ee
we find only three nontrivial gauge choices that satisfy the above conditions. We thus identify three Eulerian gauges:
\begin{itemize}
 \item[(1.)] The Newtonian/longitudinal (NL) gauge \cite{Bardeen:1980kt,Mukhanov:1990me} with: \\ 
   $A \neq 0$, $B \neq 0$, $w =0$, and $h=0$.
 
 \item[(2.)] The spatially flat (SF) gauge \cite{Kodama:1985bj,Malik:2008im,Flender:2012nq} with: \\
   $A \neq 0$, $B=0$, $w \neq 0$, and $h=0$.
 \item[(3.)] The synchronous-shear (SS) gauge with: \\
   $A =0$, $B\neq 0$, $w \neq 0$, and $h=0$.
\end{itemize}
Here, we summarise our findings for the perturbations at first order and discuss them briefly. As mentioned above the spatial displacement field is for all of these gauges the Zel'dovich displacement field, ${F}_i^{(1)}= \frac 3 2 at_0^2 \Phi_{,i}$, which immediately fixes $\dd x_i / \dd \tau \equiv \dd F_i/ \dd \tau$ in the Euler--Lagrange equation~(\ref{no_gauge}) as well.

\noindent (1.) \emph{Newtonian/longitudinal gauge.}   The perturbations in the NL gauge read
  $A_{\rm NL} = B_{\rm NL} = - \Phi$, and the temporal part of the 4-displacement field is 
$L_{\rm NL} = \Phi t$. As above, $L_{\rm NL}$ is the velocity potential of the fluid element, and thus yields a simple physical interpretation of the time-part of the 4-displacement field.
 Since we have $w_{\rm NL}=0$ in the NL gauge, the Euler--Lagrange equation~(\ref{no_gauge}) yields the Euler equation~(\ref{Euler}), where the cosmological potential on the RHS is solely given by the time perturbation $A_{\rm NL} \equiv -\Phi$. Note that the Poisson gauge reduces to the Newtonian gauge in the scalar sector.  \\
\noindent (2.) \emph{Spatially flat gauge.} The SF gauge was recently discussed in Refs.~\cite{Flender:2012nq,FlenderMaster} (sometimes called spatially Euclidean gauge) and is in particular interesting since it does not contain any perturbations in the space-space part of the metric. Thus, the 3-geometry appears Euclidean. The nonzero perturbations are
$A_{\rm SF} = -5/2 \Phi$, $w_{\rm SF}= 3/(2a) \Phi \tau$, and the temporal perturbation is 
$L_{\rm SF} = 5/2 \Phi t$. Plugging these values into the Euler--Lagrange equation~(\ref{no_gauge}), we realise that $A_{\rm SF}$ not entirely the cosmological potential but
the combination $A_{\rm SF} + \frac{\dd}{\dd \tau} a w_{\rm SF} \equiv - \Phi$. Thus, the fluid elements still move according to Newton's law of motion, but the cosmological potential receives a nonzero contribution from $w_{\rm SF}$. This feature generally 
complicates the latter physical interpretation  because $w_{\rm SF}$ sources (already at linear order) a perturbation in the expansion rate; additionally,  $w_{\rm SF}$ sources the shear as well.\footnote{The expansion rate and the shear can be defined as the trace and the traceless parts of the extrinsic curvature, respectively 
\cite{Bardeen:1980kt,Flender:2012nq}.} \\
\noindent (3.) \emph{Synchronous-shear gauge.} In contrast to the SF gauge, where the perturbations in the space-space part of the metric are zero, the perturbations in the SS gauge are only zero in the temporal part of the metric. The nonzero perturbations read 
$B_{\rm SS} = -5/3 \Phi + 2L(\fett{x})/(3\tau)$, 
$w_{\rm SS} = [L(\fett{x})- \Phi \tau]/a$, and 
$L_{\rm SS} \equiv L(\fett{x})$ is constant in time. The SS gauge has therefore a residual gauge freedom. In Ref.~\cite{Flender:2012nq}, the constant $L_{\rm SS}$ was fixed such that the density and velocity matched exactly results from Newton theory at linear order; they called this specific choice the Newtonian matter gauge. In Ref.~\cite{BruniHidalgo2013} the constant $L_{\rm SS}$ was set to zero, and they called it the Eulerian gauge.
Independently of the specific choice of $L_{\rm SS}$, the Euler--Lagrange equation~(\ref{no_gauge}) yields the Euler equation within the SS gauge, where the cosmological potential is
entirely given in terms of $\frac{\dd}{\dd \tau} a w_{\rm SS} \equiv - \Phi$.
Similarly to the SF gauge,  the SS gauge is flawed with difficulties in the physical interpretation since the nonzero $w_{\rm SS}$ distorts the Hubble diagrams and also sources cosmic shear. For recent discussions about such issues, see 
Refs.~\cite{Flender:2012nq,Haugg:2012ng}.

%%%%%%%%%%%%%%%%%%%%%%%%%%%%%%%%%%%%%%%%%%%%%%%%%%%%%%%%%%
\section{Summary and conclusions}\label{sec:con}

We find a Eulerian--Lagrangian correspondence within general relativity, which can be used to study the evolution of scalar, vector and tensor perturbations beyond leading order.
%which can be used e.g.~to study the density and velocity field in $N$-body simulations.
We restrict our analysis to an Einstein--de Sitter (EdS) Universe, although our results should  approximately hold for a $\Lambda$CDM Universe as well (see, e.g.,~\cite{Rampf:2012pu}).
Furthermore, we neglect all secondary distortions resulting from 
photons that propagate in a clumpy and expanding Universe, and we neglect biasing effects. These restrictions are obviously not realistic for cosmological observations, and it is because of this that our results at this stage can be only applied to studies on how to interpret (or setup \mbox{[quasi-]} relativistic) $N$-body simulations. We think, however, that our findings could also serve as the starting point on how to interpret cosmological observations (e.g., tracers of the velocity/density field) beyond leading order in GR.

We identify the relativistic displacement field $F^\mu(t,\fett{q})$ in terms of the coordinate/gauge transformation 
\be \label{gaugeTrafo}
x^\mu(t,\fett{q})=q^\mu+F^\mu(t,\fett{q})\,,
\ee
where $x^\mu \dot= \left(\, \tau(t,\fett{q}),\fett{x}(t,\fett{q})\,\right)$ are "some'' Eulerian coordinates (see the following) and $q^\mu \dot= (t,\fett{q})$ are
the Lagrangian coordinates. Note that the Lagrangian $\fett{q}$ ($\equiv const.$) labels the initial position of an individual fluid element.
Our starting point, see Sec.~\ref{sec:grad}, is the second-order synchronous/comoving metric with the line element
\be
  \dd s^2 = - \dd t^2 + a^2(t)\,\gamma_{ij}(t,\fett{q}) \, \dd q^i \dd q^j \,.
\ee
The 3-metric $\gamma_{ij}(t,\fett{q})$ is given in Eq.\,(\ref{gammaUniverseGrad}), and describes the gravitational evolution of an irrotational dust component in an EdS Universe. The reported synchronous metric can be obtained e.g.~from the gradient expansion technique \cite{Rampf:2012pu,Rampf:2013ewa} or from the tetrad formalism \cite{Russ:1995eu,Buchert:2013qma}. 

We then first consider a specific coordinate transformation (step by step, in Secs.~\ref{sec:GCT}--\ref{sec:sol2}), 
where the above coordinates $x^\mu \dot= (\tau,\fett{x})$ denote a Poissonian coordinate system with line element
\begin{align} \label{PoissonianSummary}
\dd s^2 &= - \big[ 1 \big. +2 \big. A \big] \dd \tau^2  
 + 2a w_i \, \dd \tau \dd x^i  \nonumber \\ 
 &\qquad\quad + a^2  \left\{ \left[ 1- 2  B \right] \delta_{ij} + S_{ij} \right\} \dd x^i \dd x^j\,, 
\end{align}
where the resulting perturbations $A$, $B$, $\fett{w}$ and $S_{ij}$ can be found in Eqs.\,(\ref{AB})--(\ref{PhiN}).
The  4-displacement field is
\be
 \label{Fmu2.}  F^\mu (t,\fett{q}) = \begin{pmatrix} L^{(1)} +L^{(2)} \\ \fett{F}^{(1)}+\fett{F}^{(2)}\end{pmatrix} \,,
\ee
where the respective quantities on the RHS can be found in Eqs.\,(\ref{firstorder_displacement}) and~(\ref{Fmu2}). 
 In the Poissonian coordinate system we identify the weak-field limit for the cosmological potential, and the occurrence of known results from Newtonian Eulerian/standard perturbation theory up to second order (cf.~the Appendix) plus relativistic corrections, where the latter become only important at scales close to the horizon. The spatial part of $F^\mu$ is the displacement field from the Newtonian LPT plus additional relativistic corrections, which again do affect the trajectories only on scales close to the horizon (for recent discussions, see also Refs.~\cite{Rampf:2012pu,Rampf:2013ewa}). 
We also find 
a transverse part in the spatial displacement field~(\ref{2LPTtrans}) which does not have any Newtonian counterpart (at that order). 
The temporal part of $F^\mu$ is the velocity potential of the fluid element 
from the Newtonian LPT plus additional relativistic corrections. 
% Note that the identification of $F^0$ being the velocity potential is only true in case of the coord trafo: synch/comoving -> poisson

Since we identify known results from Eulerian perturbation theory in the Poissonian coordinate system %(i.e., the scalar perturbations $A$ and $B$) 
and since we can relate these results to the synchronous-comoving coordinate system via the Lagrangian displacement field,
we conclude that the Poissonian coordinate system can be associated with a Eulerian frame of reference. This has two important consequences. First, 
the density and velocity in the Poissonian coordinate system have a physical significance
in the sense, that the gauge-dependent nature of the density and velocity can be associated with their frame-dependent origins. Stated in another way, since we are able to identify the Poissonian coordinate system with a Eulerian frame of reference, we deduce that the relativistic corrections of the density and velocity are not gauge artifacts. These corrections are real and thus 
measurable for a hypothetical observer who is in the Eulerian frame at rest.\footnote{Note that a hypothetical observer who is at rest in the Lagrangian frame cannot measure the \emph{Eulerian} density/velocity field. He/she only experiences/measures the \emph{Lagrangian} density/velocity change along the trajectory of a fluid element, which is labelled with initial Lagrangian coordinate $\fett{q}=const$.}
 Second, 
our results indicate that the generator of the above coordinate transformation~(\ref{gaugeTrafo}) has a direct physical significance; i.e., the generator of the coordinate transformation is the 4-displacement field, and $q^\mu+F^\mu$ is the 4-trajectory field of the fluid element with Lagrangian coordinate $\fett{q}$.
The Lagrangian and Eulerian frames are separated in terms of the displacement field, and these frames move apart from each other according to the fluid's 4-velocity, which is given in terms of the time derivative of the 4-trajectory field.
%\footnote{The fluid's velocity itself is parametrised in terms of the time derivative of the 3-displacement field.} 
The reported transverse part in the spatial displacement field yields a nonlinear frame dragging as seen in the Eulerian frame, since the transverse displacement field
 sources the frame-dragging vector potential $\fett{w}$ in the Poissonian coordinate system~(\ref{PoissonianSummary}).

Our results can be directly incorporated in Newtonian $N$-body simulations. The
 reported relativistic corrections appear as nonlinear constraints that influence the (particle's) trajectory at any time during the gravitational evolution. Since these relativistic corrections are  small with 
respect to the Newtonian bulk part, we think that the Newtonian approximation should be sufficient to model weakly nonlinear scales. However, the relativistic corrections influence (the initial
statistics of) the density and velocity field especially at scales close to the horizon.  
Thus, the relativistic corrections should be included for generating initial conditions of Newtonian $N$-body simulations, preferably in terms of
the relativistic displacement field as suggested here. Explicitly, the CDM particles are displaced from their initial grid positions according to the spatial displacement field $\fett{F}(\tau,\fett{q})$ (note that we use the Eulerian time $\tau$ to account for the initial time on the numerical grid \cite{Rampf:2013ewa}). Similarly, the peculiar velocity of the CDM particle at initial time is given by
$\fett{u}(\tau,\fett{q}) = a(\tau) \partial \fett{F}(\tau,\fett{q}) / {\partial \tau}$,
and $\fett{F}$ contains the aforementioned longitudinal and transverse components. 
The transverse displacement field does not affect the (initial) density field at second order but does affect the (initial) velocity field.
Physically, the transverse displacement field corrects for the direction of motion of the CDM particle, and neglecting it would formally yield wrong (initial and late-time) statistics for the velocity field. Technically, its practical implementation for $N$-body simulations is straightforward, and existing schemes just have to be complemented; explicit recipes to obtain initial displacements and velocities for $N$-body simulations can be found in  Refs.~\cite{Scoccimarro:1997gr,Rampf:2013ewa}.

Then, in Sec.~\ref{sec:constraint} we explicitly show that the relativistic corrections in the displacement field have to be already nonzero at initial time. Specifically,  we show that for realistic initial conditions, both the longitudinal and transverse parts of the relativistic displacement field yield initial non-linear constraints. These constraints are the result of the non-linear nature of the Einstein equations (see also section~\ref{sec:origin}), and therefore have to be included in the analysis. To our knowledge, these initial constraints have not yet been reported before in the literature.

In Sec.~\ref{sec:compare} we compare our results with the one from the literature \cite{Matarrese:1997ay,Takada:1997bk}.
Despite the fact that entirely different techniques were used, we find agreement with our results where possible. Formally, the transverse displacement field has been derived (but not identified) in terms of a gauge transformation with the use of the CPT in Ref.~\cite{Matarrese:1997ay} (note that they use the conformal time). In Ref.~\cite{Takada:1997bk}, the authors developed a post-Newtonian approach to the LPT, and also obtained the transverse displacement. Because of the complexity of their expression, we cannot confirm whether their longitudinal displacement agrees with ours. We also discuss some shortcomings related to the initial conditions for Ref.~\cite{Takada:1997bk} and about the physical interpretation for Ref.~\cite{Matarrese:1997ay}.

Finally, in Sec.~\ref{other?}, we  formulate a procedure to find all possible Eulerian gauges. For simplicity, we restrict in this part of our analysis to the scalar sector at linear order, and we shall generalise our findings in a forthcoming project. 
We find that only three gauges yield Newtonian-like trajectories together with the Zel'dovich displacement field (i.e., the weak field limit for the Eulerian and Lagrangian frames). These Eulerian gauges are (1) the Newtonian/longitudinal gauge \cite{Bardeen:1980kt,Mukhanov:1990me} which corresponds to the scalar sector of the Poisson gauge, (2) the spatially flat gauge \cite{Kodama:1985bj,Malik:2008im,Flender:2012nq}, and (3) the synchronous-shear gauge. We argue that option (1) is preferred since it comes with the easiest interpretation. Options (2) and (3), on the other hand, induce nontrivial perturbations in the trace part and the traceless part of the extrinsic curvature, and thus yield significant distortions to the Hubble diagrams and to the shear, respectively. 
Phenomenologically, such dominant distortions to the Hubble diagrams can be associated with the gravitational lensing \cite{Haugg:2012ng}; hence, options (2) and (3) might be preferred gauge choices in investigations that involve ray-tracing techniques to account for the photon propagation in a clumpy and expanding Universe.
Certainly, further investigations are necessary to explain why some Eulerian gauges are more sensitive to geometrical distortions than the others. Further understanding of this issue could support current and forthcoming efforts to interpret cosmological observations beyond leading order in GR.

\section*{Acknowledgements}

C.~R.~would like to thank Marco Bruni, Carlos Hidalgo, Karim Malik, Gerasimos Rigopoulos,  Yvonne Wong, and the referee for valuable comments on the manuscript. A part of this work contributed to the doctoral dissertation of C.~R.~at RWTH Aachen University. Parts of the results reported here were finalised during the visit of C.~R.~at the University of Portsmouth.

\appendix

\section{Comparison with the Newtonian treatment}\label{app}

In this appendix (which is based on Ref.~\cite{Rampf:2013thesis}), we wish to relate our results to the Newtonian approximation.
Let $\fett{x}$ denote the comoving coordinate defined by the rescaling of the physical coordinate $\fett{r}$ by the cosmic scale factor $a(t)$ ($\equiv (t/t_0)^{2/3}$ for an EdS universe), where $t$ is the cosmic time.
The Eulerian equations of motions for  self-gravitating dust are governed by momentum conservation, mass conservation and the Poisson equation, which are, respectively,
\begin{align}
\label{Euler1grad}  &\frac{\partial}{\partial t} \left[ a(t)\, \fett{u}(t,\fett{x}) \right] 
    + \left[ \fett{u}(t,\fett{x}) \cdot \fett{\fett{\nabla}_{\fett{x}}} \right]\, \fett{u}(t,\fett{x}) =
    - \fett{\fett{\nabla}_{\fett{x}}} \phi(t,\fett{x}) \,, \\
\label{Euler2grad} &a(t) \frac{\partial \delta(t,\fett{x})}{\partial t} 
   + \fett{\fett{\nabla}_{\fett{x}}} \cdot \left\{ \left[ 1+ \delta(t,\fett{x}) \right] \fett{u}(t,\fett{x}) \right\} =0\,, \\
\label{Euler3grad} &\fett{\nabla}_{\fett{x}}^2 \phi(t,\fett{x}) = \frac 3 2 H^2(t) \,a^2(t) \,\delta(t,\fett{x}) \,, 
\end{align}
where $\fett{u} \!=\! a\, \partial \fett{x} / \partial t$ is the peculiar velocity of the fluid particle, $H=2/(3t)$ for an EdS 
universe, $\phi$ is the cosmological potential, and the density contrast $\delta(t,\fett{x})$ separates the local variation of the mass density $\rho(t,\fett{x})$ from a global background $\overline\rho(t)$: $\rho(t,\fett{x}) = \overline\rho(t) [1+\delta(t,\fett{x})]$. 
Furthermore, we demand an irrotational fluid motion: $\fett{\fett{\nabla}_{\fett{x}}} \times \fett{u} =\fett{0}$.

A convenient way to solve the above set of equations is to use the Newtonian LPT (e.g., Refs.~\cite{Buchert:1992ya,Bouchet:1994xp,BuchertRampf:2012} and references in Ref.~\cite{Bernardeau:2001qr}). In the Newtonian LPT, the observer follows the trajectories of the individual fluid elements, where the dynamical information of the trajectory field is encoded in the displacement field $\fett{\Psi}$. (To avoid confusion with the relativistic displacement field, we label the Newtonian one with $\fett{\Psi}$ instead of $\fett{F}$). The coordinate mapping from the fluid particles' initial position $\fett{q}$ plus its gravitationally induced displacement is then given by
\be
  \fett{x}(t) = \fett{q} + \fett{\Psi}(t,\fett{q}) \,.
\ee
The displacement field contains all the dynamical information of the system, and the fluid displacement automatically obeys mass conservation by the relation
\be
\label{massGrad} \delta(t,\fett{x}) = \frac{1}{\det[\delta_{ij} + \Psi_{i,j}]} -1 \,,
\ee
with the Jacobian of the transformation $J = \det[\delta_{ij} + \Psi_{i,j}]$, where ``$,j$'' denotes a spatial differentiation 
w.r.t.~Lagrangian coordinate $q_j$, and $i,j,\ldots = 1 \ldots 3$. In the LPT the above relation replaces the mass 
conservation~(\ref{Euler2grad}), where the neglect of an integration constant $\delta_0$ can always be justified in the Newtonian limit, i.e., by a proper set of initial conditions,  by using a different set of Lagrangian coordinates, or by the assumption of an initial quasihomogeneity; see  Ref.~\cite{BuchertRampf:2012}.

In the Newtonian LPT, the system~(\ref{Euler1grad})--(\ref{Euler3grad}), together with
the irrotationality constraint is solved with a perturbative ansatz for the displacement field $\fett{\Psi}$,
 which is supposed to be a small quantity:
\be
 \label{ansatzLPTGrad}  \fett{\Psi}(t,\fett{q}) = \sum_{i=1}^\infty\fett{\Psi}^{(i)}(t,\fett{q}) \,.
\ee
Usually, one utilises the Newtonian LPT within a restricted class of initial conditions where only one initial piece of data has to be given \cite{Buchert:1992ya} 
(this class is of the Zel'dovich type \cite{Zeldovich:1969sb}). Then, the 
initial data at time $t_0$ is given by the initial gravitational potential $\Phi(t_0,\fett{q})$ (up to some arbitrary constants) only, 
which is supposed to be smooth and of order $10^{-5}$. Solving the above in NLPT up to second order one finds for the fastest growing 
solutions \cite{BuchertRampf:2012},
\begin{align}
 \label{2LPTGrad}  \Psi_i(t,\fett{q}) =  &\left( \frac 3 2 \right) a(t) \,t_0^2\, \Phi_{,i}(t_0,\fett{q}) \nonumber \\ 
&\quad -\left( \frac 3 2 \right)^2 \frac 3 7 a^2(t) \, t_0^4  \frac{\partial}{\partial q_i} \frac{1}{\fett{\nabla}_{\fett{q}}^2} \mu_2(t_0,\fett{q}) + {\cal O}(\Phi^3) \,, 
\end{align}
where $1/\fett{\nabla}_{\fett{q}}^2$ is the inverse Laplacian and $\mu_2(t_0,\fett{q})=1/2(\Phi_{,ll}\Phi_{,mm}-\Phi_{,lm}\Phi_{,lm})$. 
%In the following it is sufficient to truncate the Jacobian up to two potentials $\Phi$ only:
%\be
%  J = 1 + left( \frac 3 2 \right) a(t) \,t_0^2 \Phi_{,ll} + \frac 9 7 a^2(t) t_0^2 \mu_2  + {\cal O}(\Phi^3) \,.
%\ee
Now, what is the effect on the Poisson equation; specifically, what is the relation between the cosmological potential $\phi(t,\fett{x})$ 
and the initial gravitational potential $\Phi$? To see this, we plug Eq.~(\ref{massGrad}) into the Poisson equation~(\ref{Euler3grad}), i.e., 
\begin{align}
  \fett{\nabla}_{\fett{x}}^2 \phi(t,\fett{x}) &= \frac 2 3 \frac{a^2(t)}{t^2} \left(  \frac{1}{\det[\delta_{ij} + \Psi_{i,j}]} -1
  \right) \,,
\intertext{and with the use of the second-order displacement field~(\ref{2LPTGrad}) we Taylor expand the RHS. Then we obtain}
  \label{PoissonExpanded} \fett{\nabla}_{\fett{x}}^2 \phi(t,\fett{x}) &= - \Phi_{,ll}(t_0,\fett{q}) -\frac 6 7 a(t) \, t_0^2 \, \mu_2(t_0,\fett{q})  \nonumber \\
   & + \frac 3 2 a(t)\, t_0^2\, \Phi_{,ll} (t_0,\fett{q}) \, \Phi_{,mm}(t_0,\fett{q}) + {\cal O}(\Phi^3) \,.
\end{align}
{Note that the LHS is a Eulerian quantity, whereas the expressions on the RHS depend on Lagrangian coordinates and Lagrangian derivatives. We expand the dependences and interchange the derivatives (we denote ``$|_i$'' for the differentiation w.r.t. Eulerian coordinate $x_i$) on the RHS and finally multiply the whole equation with a $1/\fett{\nabla}_{\fett{x}}^2$. Then, we have} 
\begin{align}
  \label{gravPotGrad} \phi(t,\fett{x}) &= - \Phi(t_0,\fett{x}) +\frac 3 4 a(t) \, t_0^2 \, \Phi_{|l}(t_0,\fett{x}) \, \Phi_{|l}(t_0,\fett{x}) 
\nonumber \\    &\qquad 
  +\frac{15}{7} a(t)\, t_0^2 \frac{1}{\fett{\nabla}_{\fett{x}}^2} \overline{\mu}_2(t_0,\fett{x}) \,,
\end{align}
{with $\overline{\mu}_2(t_0,\fett{x})$ analogous to ${\mu}_2(t_0,\fett{q})$ but the dependences and derivatives are w.r.t.~$\fett{x}$.
The above has been obtained in Ref.~\cite{Stewart:1994wq} (though their approach differs from ours; also, cf.~the first bracketed term of Eq.~(6.8) in \cite{Matarrese:1997ay}). To see its connection to the ``Newtonian literature'',
we expand the second term on the RHS with $\fett{\nabla}_{\fett{x}}^2 /{\fett{\nabla}_{\fett{x}}^2}$, which leads to}
\begin{align}
  \label{gravPotNewt} \phi(t,\fett{x}) &= - \Phi(t_0,\fett{x}) +\frac 3 2 a(t) \, t_0^2  \,  \frac{1}{\fett{\nabla}_{\fett{x}}^2} F_2(t_0,\fett{x}) \,,
\end{align}
{where we have defined}
\begin{align}
 \label{F2Grad}  F_2(t_0,\fett{x})  &= \frac 5 7 \Phi_{|ll} \, \Phi_{|mm}  + \Phi_{|l} \, \Phi_{|lmm}  %\Bigg. \nonumber \\ 
  %  &\qquad\qquad\qquad\quad\Bigg.
+ \frac 2 7 \Phi_{|lm} \, \Phi_{|lm} \,.
\end{align}
This is nothing but the result expected from standard perturbation theory (SPT) up to second order (see, e.g., Eq.~(45) in 
Ref.~\cite{Bernardeau:2001qr}). Equation~(\ref{gravPotGrad}) or Eq.~(\ref{gravPotNewt}) can be interpreted as follows. 
At leading order, the cosmological potential is just proportional to the initial gravitational potential, whereas at second order, the temporal 
extrapolation of the initial tidal field leads to an ``evolving'' cosmological potential. Note that 
expression~(\ref{gravPotNewt}) is identical with~(\ref{PhiN}), where the latter was obtained in the relativistic coordinate transformation~(\ref{gct}).

Similar considerations can be made for the peculiar fluid velocity. 
We connect the fluid velocity to the initial gravitational potential. Up to second order in the conventional Newtonian LPT the fluid 
motion is purely potential in the Lagrangian frame \cite{Buchert:1993ud,Bouchet:1994xp}, so we are allowed to introduce a (peculiar) velocity potential ${\cal S}$ such that 
\be
\fett{u}(t,\fett{x}) =  \frac{\fett{\nabla_x} {\cal S}(t,\fett{x})}{a(t)} \equiv \fett{\nabla_r} {\cal S} \,, 
\ee
 and plug it into the Euler equation~(\ref{Euler1grad}). The very equation can then
be integrated w.r.t.~$\fett{x}$ and it yields  the Bernoulli equation 
\cite{Kofman:1988ak,KofmanPogosyan,Jones1999}
(it is equivalent to the non-relativistic Hamilton--Jacobi equation; see, e.g., Ref.~\cite{Stewart:1994wq})
\begin{align}
\label{Bernoulli} \frac{\partial}{\partial t} {\cal S}(t,\fett{x}) +\frac{1}{2a^2(t)} 
    \left[\fett{\nabla}_{\fett{x}} {\cal S}(t,\fett{x}) \right]^2 = - \phi(t,\fett{x}) \,,
\end{align}
where $\phi$ is explicitly given in Eq.~(\ref{gravPotGrad}) up to second order. 
Here, we have set an integration constant $c(t)$ to zero since it can always be absorbed into the velocity potential by replacing 
${\cal S} \rightarrow  {\cal S} + \int c(t)\, {\rm d} t$; so it does not affect the flow \cite{Jones1999}.

We solve the above differential equation with a recursive technique, assuming the usual series hierarchy within the SPT. Then, we obtain for the peculiar-velocity potential
\begin{align}
 {\cal S}(t,\fett{x}) &= \Phi(t_0,\fett{x}) \, t -\frac 3 4 t_0^{4/3} t^{5/3} \, \Phi_{|l} \, \Phi_{|l}
  -\frac 9 7 t_0^{4/3} t^{5/3} \frac{1}{\fett{\nabla}_{\fett{x}}^2} \overline{\mu}_2 \nonumber \\ 
  &\equiv  \Phi(t_0,\fett{x}) \, t -\frac 3 2 t_0^{4/3} t^{5/3}  \frac{1}{\fett{\nabla}_{\fett{x}}^2} G_2(t_0,\fett{x})  \,,
\end{align}
{with}
\begin{align}
 \label{G2Grad}  G_2(t_0,\fett{x}) &=  \frac 3 7 \Phi_{|ll}  \, \Phi_{|mm}
    + \Phi_{|l}  \, \Phi_{|lmm} %\Bigg. \nonumber \\
 % &\qquad\qquad \Bigg.
 + \frac 4 7 \Phi_{|lm} \, \Phi_{|lm}   \,,
\end{align}
{or interchanging the dependences and derivatives to be Lagrangian}
\begin{align}
\label{velPotGrad}  {\cal S}(t,\fett{q}) &= \Phi(t_0,\fett{q}) \, t +\frac 3 4 t_0^{4/3} t^{5/3} \, \Phi_{,l} \, \Phi_{,l}
  -\frac 9 7 t_0^{4/3} t^{5/3} \frac{1}{\fett{\nabla}_{\fett{q}}^2} {\mu}_2 \,.
\end{align}
Again, this is the second-order result for the velocity potential from the SPT \cite{Bernardeau:2001qr}.
The expression~(\ref{velPotGrad}) is identical with the nonrelativistic part in the time perturbation $L$; see Eq.\,(\ref{eqL2}).

In summary, we have calculated the nonrelativistic perturbations $\phi$ and $\cal S$, which agree exactly with their counterparts in the Poissonian metric (see Sec.~\ref{sec:con}).

%%%%%%%%%%%%%%%%%%%%%%%%%%%%%%%%%%%%%%%%%%%%%%%%%%%%%%%%

%%%%%%%%%%%%%%%%%%%%%%%%%%%%%%%%%%%%%%%%%%%%%%%%%%%%%%%

\end{document}